\documentclass[12pt]{iopart}
\usepackage{amssymb}
\usepackage{iopams}
\expandafter\let\csname equation*\endcsname\relax

\expandafter\let\csname endequation*\endcsname\relax
\usepackage{amsmath}
\usepackage{graphicx}
\usepackage{bm}
\usepackage{braket}
\usepackage{color}
\begin{document}
\title[Inelastic scattering of vortex electrons]{Inelastic scattering of vortex electrons beyond the Born approximation}
\author{S Strnat$^{1, 2}$, J Sommerfeldt$^{3}$, A K Sahoo$^4$,
 L Sharma$^4$ and A Surzhykov$^{1, 2}$}
\address{$^1$ Fundamentale Physik für Metrologie FPM, Physikalisch-Technische Bundesanstalt PTB, D-38116 Braunschweig, Germany}
\address{$^2$ Institut für Mathematische Physik, Technische Universität Braunschweig, D-38106 Braunschweig, Germany}
\address{$^3$ Laboratoire Kastler Brossel, Sorbonne Université, CNRS, ENS-Université PSL, Collège de France, Campus Pierre et Marie Curie, Paris, France}
\address{$^4$ Indian Institute of Technology Roorkee, Roorkee 247667, India}

\ead{sophia.strnat@ptb.de}
\begin{abstract}
We present a theoretical study of the inelastic scattering of vortex electrons by a hydrogen atom. In our study, special emphasis is placed on the effects of the Coulomb interaction between a projectile electron and a target atom. To understand these effects, we construct vortex electron wave functions both from free space and distorted solutions of the Schrödinger equation. These wave functions give rise to the first Born and distorted wave scattering amplitudes, respectively. The derived theory has been employed to investigate the $1s \rightarrow 2p$ transition of a hydrogen atom induced by electrons with the kinetic energies in the range from 20 to 100 eV. The results of the calculations have clearly indicated that the Coulomb interaction can significantly affect the phase pattern and probability density of a vortex electron beam as well as the squared transition amplitudes. For the latter, the most pronounced effect was found for the excitation to the $\ket{2p\, m_f=0}$ sublevel and large scattering angles.
\end{abstract}
\section{Introduction}
Vortex (also referred to as "twisted") electrons represent the cylindrical solutions of the Schrödinger (or Dirac) equation and are characterized by a non-zero orbital angular momentum projection along their propagation direction, a helical phase front, and an annular probability density profile in the transverse plane~\cite{bliokh2007semiclassical, BlIv17, lloyd2017electron}. Since the first experimental realization in 2010 \cite{uchida2010generation, verbeeck2010production}, there has been significant interest in their potential applications. In particular, vortex electrons were found to be a powerful tool in electron microscopy where they help in determining magnetic states of materials~\cite{verbeeck2010production} and chirality of crystals \cite{juchtmans2015using, juchtmans2015orbital} or investigating the local properties of nanomaterials and biomolecules \cite{asenjo2014dichroism}. A deep theoretical understanding of the underlying fundamental processes and, first of all, the electron-atom scattering is highly demanded for these applications. In the last decade, therefore, a number of theoretical investigations have been presented to discuss the elastic and inelastic scattering of vortex electrons by atomic targets.

Most of the investigations so far have dealt with elastic electron-atom scattering~\cite{edstrom2016elastic, van2014rutherford, juchtmans2015using, maiorova2018elastic, serbo2015scattering}. Beside the first Born approximation, this scattering has been evaluated also within the distorted wave theory where the interaction between the projectile electrons and an atom has been taken into account more accurately~\cite{kosheleva2018elastic, Ivanov:2022jzh, harris2024distorted}. Comparisons between first Born and distorted wave calculations have stressed the importance of a proper account of the Coulomb interaction between a projectile electron and the target atom that leads to significant corrections to the total and differential cross sections. Much less attention has been paid so far to the inelastic scattering of vortex electrons, leading to the excitation of a target atom. The first step towards such an analysis has been performed within the framework of non-relativistic first Born approximation by van Boxem and coauthors in Ref.~\cite{van2015inelastic}. The first Born approach is expected to be well-justified for the high-energy regime of about 100 keV and above but should be questioned for lower energies. The latter domain attracts currently particular attention because of potential use of vortex beams in low energy electron microscopes~\cite{HAWKES2014110} or by producing them with the Kapitza-Dirac effect~\cite{Handali:15}. These studies deal with electron energies from few eV up to few hundred eV for which the interaction with the Coulomb center can be of particular importance~\cite{bauer1994low, Handali:15, Chauhan2005}.

In this contribution, we present a theoretical investigation of the inelastic scattering of vortex electrons by a hydrogen target atom. To better understand and illustrate the effect of the Coulomb distortion, we start with the analysis of the wave functions of the electrons propagating both in free space and in the presence of a Coulomb center in Sec.~\ref{sec:wf}. With the help of these wave functions we derive in Sec.~\ref{sec:scatamp} the amplitude of the inelastic electron scattering within the first Born and distorted wave approximations. The numerical details needed for the computation of these amplitudes are given briefly in Sec.~\ref{sec:numerics}. The results of the calculations of both the projectile wave functions and the scattering amplitudes are presented and discussed in detail in Sec.~\ref{sec:results}. Here in particular, we demonstrate that the interaction with the Coulomb center can significantly affect the phase front and the probability density profile of the electron beam. Also the scattering amplitude and, hence, the cross section can be also influenced by the Coulomb distortion and the effect becomes most pronounced for the higher scattering angles. The conclusion is given finally in Sec.~\ref{sec:sum}.\\
Atomic units ($m = 1$, $e =1$ and $\hbar = 1$) are used throughout this article unless stated otherwise.
\section{Geometry and basic notation}\label{sec:geometry}
\begin{figure}
    \centering
    \includegraphics[width=0.75\linewidth]{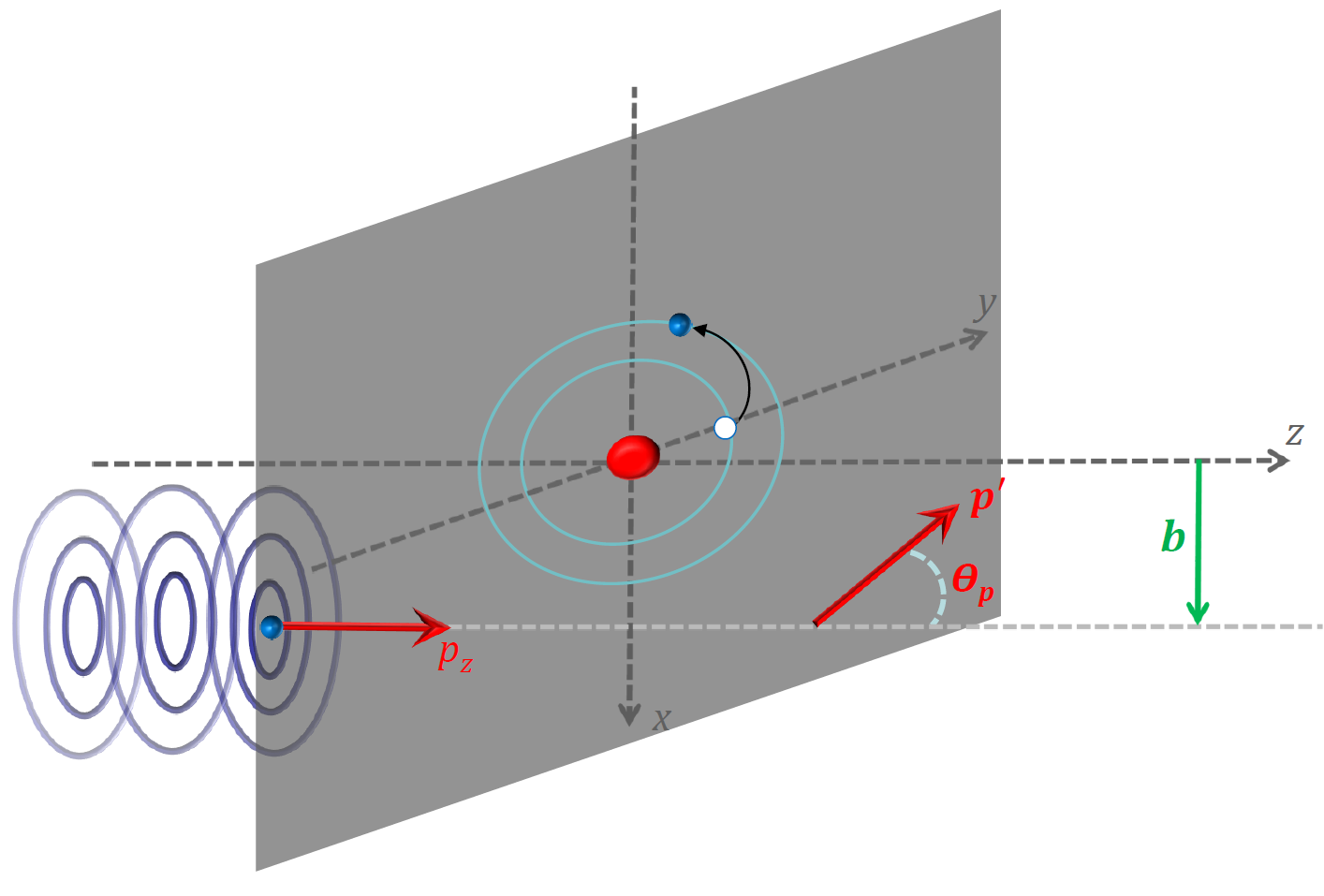}
    \caption{The geometry of inelastic scattering of a vortex electron beam by a target hydrogen atom is considered. The incident electron, which is assumed to be in the Bessel state with well-defined longitudinal momentum $p_z$ and projection of orbital angular momentum $m_\ell$ on the propagation axis. This axis is shifted from the coordinate origin by the vector $\bm b$.}
    \label{geometry}
\end{figure}

Before we discuss the theory of inelastic scattering of vortex electrons, we shall first consider the geometry of the process and its basic parameters. As shown in Fig. \ref{geometry}, a target atom is placed at the origin of the coordinate system and undergoes a $1s \rightarrow 2p$ transition during the scattering. The incident vortex electron beam propagates parallel to the $z$ axis and has a well-defined longitudinal momentum $p_z$. The center of the incoming beam, also known as the vortex line, is displaced by the impact parameter $\bm b$ from the $z$ axis. After scattering, the electron is assumed to be detected in a non-vortex state, which possesses a well-defined asymptotic momentum $\bm p'$. This momentum is tilted with respect to the $z$ axis by the polar angle $\theta_{\bm p'}$ and lies in the plane defined by the atom and the vortex line of the incident beam.

\section{Electron wave functions}\label{sec:wf}
The analysis of inelastic scattering of vortex electrons by an atomic target requires first constructing the incident electron wave function. This wave function should account for both the vorticity of an electron and the distortion caused by the Coulomb field of a target atom. To demonstrate how such a wave function can be derived and to discuss its basic properties, we begin in Sec.~\ref{conv_solutions} by examining the free and distorted non-vortex solutions of the Schrödinger Hamiltonian. Later, in Sec.~\ref{subsec_ve}, we will proceed with the construction of vortex electron states.
\subsection{Non-vortex electrons: Free and distorted solutions}\label{conv_solutions}
Below, we will consider an unbound solution of the non-relativistic Schrödinger Hamiltonian
\begin{equation}\label{hamiltonian}
    \hat{H} = \frac{\hat{\bm p}^2}{2} + V (\bm r),
\end{equation}
where $\hat{\bm p}$ is the linear momentum operator, and $V(\bm r)$ describes the interaction of an electron with a scattering center. In the case where this interaction can be neglected, i.e. $V(\bm r) = 0$, the plane waves
\begin{equation}\label{planewave}
    \psi_{\bm p} (\bm r) = e^{i \bm p \cdot \bm r},
\end{equation}
with momentum vector $\bm p$ and coordinate vector $\bm r$, are the well-known solutions of Eq.~\eqref{hamiltonian}. For the theoretical analysis below, it may be convenient to expand the plane wave~\eqref{planewave} into partial waves:
\begin{equation}\label{pwex_pw}
    \psi_{\bm p} (\bm r) = 4 \pi \sum_{l=0}^\infty \sum_{m=-l}^l i^l j_l(p r) Y_{l m}(\hat{\bm r}) Y_{l m}^*(\hat{\bm p}),
\end{equation}
where $j_l(p r)$ are the spherical Bessel functions, and $Y_{l m}$ are the spherical harmonics. Here, $\hat{\bm r} \equiv (\theta_{\bm r}, \varphi_{\bm r})$ and $\hat{\bm p} \equiv (\theta_{\bm p}, \varphi_{\bm p})$ define the directions of the coordinate and momentum vector, respectively \cite{arfken2011mathematical}.

For an accurate treatment of the scattering processes, the wave functions should account for the interaction between the incoming electron and the scattering center. This requires a solution of the Schrödinger equation with the Hamiltonian~\eqref{hamiltonian} in which $V(\bm r) \neq 0$. By assuming the interaction potential to be spherically symmetric, $V(\bm r) = V(r)$, this solution can be written in general as
\begin{equation}\label{dw_wf_eq}
    F_{\bm p}^{(\pm)}(\bm r) =\frac{4 \pi}{p r} \sum_{l = 0}^\infty \sum_{m= -l}^{l} i^l \phi_l^{(\pm)}(p, r) Y_{lm}(\hat{\bm r}) Y_{lm}^*(\hat{\bm p}),
\end{equation}
which closely resembles the partial wave expansion of the plane wave~\eqref{pwex_pw}. The essential difference, however, is that the radial wave functions $\phi_l^{(\pm)}(p, r)$ are used instead of the spherical Bessel functions as in the plane wave case. The explicit form of $\phi_l^{(\pm)}(p, r)$ depends on the particular choice of the interaction potential $V(r)$ and will be discussed later in Sec.~\ref{sec:numerics}.

The wave function~\eqref{dw_wf_eq}, obtained for the non-vanishing interaction potential, is usually referred to as the distorted wave solution. In contrast to the plane wave, the asymptotic form of $F_{\bm p}^{(\pm)}(\bm r)$ for large distances from the scattering center 
\begin{equation}\label{distortedwave}
    F_{\bm p}^{(\pm)}(\bm r) \Big|_{r \rightarrow \infty} \Big.= e^{i \bm p \cdot \bm r} + f^{DW} \frac{e^{\pm i p r}}{r},
\end{equation}
can be written as a sum of a plane wave and an outgoing (incoming) spherical wave, where the latter is weighted by the scattering amplitude $f^{DW}$ \cite{burke2013theory}.
\subsection{Vortex electrons: Free and distorted solutions} \label{subsec_ve}
Having briefly recalled the well-known formulas for the conventional (non-vortex) free and distorted wave solutions of the Schrödinger equation, we are now ready to discuss their vortex counterparts. Similar to Sec.~\ref{conv_solutions}, we begin with the case of a vanishing interaction potential, $V(\bm r) = 0$. For this case, one can construct solutions of the Schrödinger equation that possess well-defined longitudinal momentum $p_z$ and projection $m_\ell$ of the orbital angular momentum (OAM) on the propagation direction. These so-called Bessel solutions can be written as a coherent superposition of plane waves:
\begin{equation}\label{pw_vortex_wf}
        \psi^{V}_{m_\ell \varkappa p_z} (\bm r) = \int \frac{d ^3\bm p}{2 \pi}\ a_{m_\ell \varkappa p_z}(\bm p)\ e^{i \bm p \cdot \bm r},
\end{equation}
whose momenta $\bm p$ lie on the surface of a cone with opening angle $\theta_{\bm p}$. Here, the amplitude 
\begin{equation}\label{amplitude_eq}
    a_{m_\ell \varkappa p_z}(\bm p) = (-i)^{m_\ell}\ e^{i m_\ell \varphi_{\bm p}}\ \frac{\delta(|\bm p_\perp| - \varkappa)}{\varkappa} \delta(p_\parallel - p_z),
\end{equation}
ensures a well-defined absolute transverse momentum $|\bm p_\perp| = \varkappa$, and $\varphi_{\bm p}$ is the azimuthal angle. By inserting this amplitude into Eq.~\eqref{pw_vortex_wf} and performing some simple algebra, we find 
\begin{equation}\label{bessel}
    \psi^{V}_{m_\ell \varkappa p_z} (\bm r) = e^{i m_\ell \varphi_{\bm r}} J_{m_\ell}(\varkappa r_\perp) e^{i p_z z},
\end{equation}
where $J_{m_\ell}(\varkappa r_\perp)$ is the Bessel function of the first kind, which gives its name to the solution. As seen from this formula, the Bessel states exhibit a characteristic annular structure in the transverse $(x y)$ plane with a dark center for $m_\ell \neq 0$. This dark center corresponds to a phase singularity and marks the position of the vortex line in the $x y$ plane (see Refs.~\cite{larocque2018twisted, BlIv17, verbeeck2012new, mcmorran2011electron} for further details).

Expressions~\eqref{pw_vortex_wf}-\eqref{bessel} are derived for the electron beam whose vortex line passes through the coordinate origin. In order to shift the vortex line by the transverse vector $\bm b = b\ (\cos{\varphi_{\bm b}}, \sin{\varphi_{\bm b}}, 0)^T$, the standard translation operator $\hat{T} = e^{-i \hat{\bm p} \cdot \bm b}$ is used. By applying this operator to the wave function $\psi^{V}_{m_\ell \varkappa p_z} (\bm r)$, we obtain
\begin{equation}
    e^{-i \hat{\bm p} \cdot \bm b} \psi^{V}_{m_\ell \varkappa p_z} (\bm r) = \psi^{V}_{m_\ell \varkappa p_z \bm b} (\bm r) =  \int \frac{d ^3\bm p}{2 \pi}\ a_{m_\ell \varkappa p_z}(\bm p)\ e^{-i \bm p \cdot \bm b} e^{i \bm p \cdot \bm r}.
\end{equation}
It is worth mentioning that this solution still possesses a well-defined OAM projection $m_\ell$ on the shifted propagation axis which no longer coincides with the $z$ axis:
\begin{equation}
    \hat{L}_{z, \bm b} \psi^{V}_{m_\ell \varkappa p_z \bm b} (\bm r) = \hat{T} \hat{L}_z \hat{T}^\dagger \hat{T} \psi^{V}_{m_\ell \varkappa p_z} (\bm r) = m_\ell \hat{T} \psi^{V}_{m_\ell \varkappa p_z} (\bm r) = m_\ell \psi^{V}_{m_\ell \varkappa p_z \bm b} (\bm r),
\end{equation}
see Fig.~\ref{geometry}. Here, $\hat{L}_{z, \bm b}$ denotes the $\hat{L}_z$ operator shifted from the coordinate origin by $\bm b$, and $\hat{T}^\dagger \hat{T} = \hat{I}$ was used.

We are prepared now to construct vortex solutions, distorted by the Coulomb interaction with the target atom. This scenario requires particular attention since the geometry of the problem is defined now by the position of the vortex line and of the Coulomb center. In the simplest case, the vortex line passes through this center which corresponds to $\bm b = (0, 0, 0)^T$ in Fig.~\ref{geometry}. For this geometry, the vortex wave function can be constructed 
\begin{equation}\label{dw_vortex_wf_b0}
    F^{V, (\pm)}_{m_\ell \varkappa p_z} (\bm r) = \int \frac{d ^3\bm p}{2 \pi}\ a_{m_\ell \varkappa p_z}(\bm p)\ F_{\bm p}^{(\pm)} (\bm r),
\end{equation}
as a coherent superposition of distorted waves~\eqref{dw_wf_eq}, where the amplitude $a_{m_\ell \varkappa p_z}(\bm p)$ is given by Eq.~\eqref{amplitude_eq}. A detailed discussion of this expression is given in Refs. \cite{ZaSe17, kosheleva2018elastic, maiorova2021radiative}.

In order to treat a more general case where the vortex line is displaced from a target atom (i. e., from the coordinate origin), a naive approach would be to apply the translation operator to the \textit{entire} distorted vortex wave function:
\begin{equation}\label{dw_vortex_wf_t}
    \begin{split}
        e^{- i \hat{\bm p} \cdot \bm b} F^{V, (\pm)}_{m_\ell \varkappa p_z} (\bm r) & = \int \frac{d ^3\bm p}{2 \pi}\ a_{m_\ell \varkappa p_z}(\bm p)\ e^{- i \hat{\bm p} \cdot \bm b} F_{\bm p}^{(\pm)} (\bm r) \\\
        & = c \int \frac{d ^3\bm p}{2 \pi}\ a_{m_\ell \varkappa p_z}(\bm p)\ e^{- i \bm p \cdot \bm b} e^{ i \bm p \cdot \bm r} e^{- i \hat{\bm p} \cdot \bm b} g^{(\pm)}(\bm r)\\\
        & = c \int \frac{d ^3\bm p}{2 \pi}\ a_{m_\ell \varkappa p_z}(\bm p)\ e^{- i \bm p \cdot \bm b} e^{ i \bm p \cdot \bm r} g^{(\pm)}( \bm r - \bm b).
    \end{split}
\end{equation}
In the second and third lines of this equation, we have used the fact that, for a spherically symmetric potential, the function $F_{\bm p}^{(\pm)} (\bm r)$ can be generally written
\begin{equation}\label{generalform_f}
    F_{\bm p}^{(\pm)} (\bm r) = c e^{i \bm p \cdot\bm r} g^{(\pm)}( \bm r),
\end{equation}
as the product of a plane wave and a function $g^{(\pm)}(\bm r)$ that describes the interaction of a projectile electron with a target. This formula is well known, for example, for a pure Coulombic potential where $g^{(\pm)}(\bm r)$ is given by the confluent hypergeometric function $_1 F_1$, see Ref.~\cite{eichler2005lectures} for further details. 

The naive approach represented by Eq.~\eqref{dw_vortex_wf_t} does not ``separate" the scattering center from the vortex line since it shifts both the plane wave and spherical wave part of the distorted wave function by $\bm b$. In order to spatially isolate these two parts from each other, one can use a simple mathematical trick. First, we introduce an additional phase factor $e^{- i \bm p \cdot \bm b}$ to Eq.~\eqref{dw_vortex_wf_t}, which ensures that the vortex line is displaced by $2 \bm b$ from the coordinate origin while the center of the spherical wave remains at $\bm b$. By performing then a general coordinate transformation $\bm r \rightarrow \bm r + \bm b$, we finally obtain
\begin{equation}\label{dw_vortex_wf_eval}
        F^{V, (\pm)}_{m_\ell \varkappa p_z \bm b} (\bm r) = \int \frac{d ^3\bm p}{2 \pi}\ a_{m_\ell \varkappa p_z}(\bm p)\ e^{-i \bm b \cdot \bm p} F_{\bm p}^{(\pm)} (\bm r).
\end{equation}
This expression describes a distorted vortex wave function constructed from solutions of the Schrödinger equation with a target atom located at the coordinate origin and the vortex at $\bm b$. A similar expression has been used to analyze the relativistic collision of vortex electrons in Refs.~\cite{ZaSe17, kosheleva2018elastic, maiorova2021radiative}.

By inserting the amplitude~\eqref{amplitude_eq} into Eq.~\eqref{dw_vortex_wf_eval} and performing some simple algebra, we find the multipole expansion of the distorted vortex wave
\begin{equation}\label{distorted_nv_wf}
        F^{V, (\pm)}_{m_\ell \varkappa p_z \bm b} (\bm r)  = (2 \pi)^\frac{3}{2} \sum_{l m} (-1)^{m_\ell-m} i^{(l-m)} \phi_l^{(\pm)}(p, r) Y_{lm} (\hat{\bm r}) J_{m_\ell -m} (\varkappa b) \Theta_{lm}^* (\theta_{\bm p}).
\end{equation}
where the radial components $\phi_l^{(\pm)}(p, r)$ are the same as in Eq.~\eqref{dw_wf_eq}. Moreover, we introduced the angular function
\begin{equation}
    \Theta_{l m} (\theta) = \sqrt{\frac{1}{2 \pi}} \sqrt{\frac{2 l + 1}{2} \frac{(l-m)!}{(l+m)!}} P_{l}^m (\cos \theta).
\end{equation}
with $P_l ^{m}$ being the associated Legendre polynomials.
%
\section{Scattering amplitudes for the $1s \rightarrow 2p$ transistion}\label{sec:scatamp}
With the help of the wave functions from Sec.~\ref{sec:wf}, we will construct now the amplitudes to describe inelastic electron scattering. To simplify our analysis, we will restrict ourselves to the $1s \rightarrow 2p$ excitation of the hydrogen atom which usually serves as a testbed for various atomic collision theories. 
Specifically, in Sec.~\ref{subsec:ba}, the well-known Born approximation will be briefly discussed both for conventional and vortex electrons. A more advanced distorted wave approach will be presented then in Sec.~\ref{subsec:dw}.
\subsection{Born approximation} \label{subsec:ba}
Not much has to be said about the first Born approach to describe inelastic scattering of plane wave electrons since it has been discussed in detail in the literature~\cite{moiseiwitsch1968electron, holt1968application}. Within this framework, the amplitude for the excitation from the $\ket{1s}$ ground state (with magnetic quantum number $m_i = 0$) to a $\ket{2p\, m_f}$ magnetic sublevel reads as follows
\begin{equation}\label{fbnonv}
    \begin{split}
    f_{2p\, m_f, 1s}^{FB} & = \frac{1}{(2 \pi)^3} \int d^3 \bm r_0\ d^3 \bm r_1 e^{i \bm q \cdot \bm r_0} \psi_{2p\, m_f}^* (\bm r_1) \, \frac{1}{| \bm r_0 - \bm r_1|} \, \psi_{1s}(\bm r_1) \\\
    & = (-1)^{-m_f} \frac{128 \sqrt{6} i}{q ( 4 q^2 + 9)^3 \pi^{3/2}}\, Y_{1 -m_f}(\hat{\bm q}) ,
    \end{split}
\end{equation}
where $1/|\bm r_0 - \bm r_1|$ is the Coulomb interaction operator with the projectile and target electron coordinates $\bm r_0$ and $\bm r_1$, and $\bm q = \bm p - \bm p'$ is the momentum transfer. Moreover, we used the explicit form of the $\ket{1s}$ and $\ket{2p\, m_f}$ wave functions 
\begin{subequations} \label{eq:wavefunctions}
    \begin{align}
        \psi_{1s}(\bm r) &= \frac{2\ e^{-r}}{\sqrt{4 \pi }} \label{eq:1a}, \\
        \psi_{2p\, m_f}(\bm r) &= \frac{r \ e^{-r/2}}{2 \sqrt{6}}\, Y_{1 m_f} (\hat{\bm r}) \label{eq:1b},
    \end{align}
\end{subequations}
the multipole expansion of the plane wave $e^{i \bm q \cdot \bm r_0}$, and the well-known Bethe integral ansatz to arrive to the second line of Eq.~\eqref{fbnonv}.

The expression~\eqref{fbnonv} has been derived for non-vortex (plane wave) scattering electron states. This Born approximation can be trivially extended to the case when the incident electron is in the Bessel state:
\begin{equation}\label{amp_pw_v}
    \begin{split}
     f_{2p\, m_f, 1s}^{FB, V} & = \frac{1}{(2 \pi)^3} \int d^3 \bm r_0\ d^3 \bm r_1 e^{-i \bm p' \bm r_0} \psi_{2p\, m_f}^* (\bm r_1) \frac{1}{| \bm r_0 - \bm r_1|} \psi_{1s}(\bm r_1)  \psi^{V}_{m_\ell \varkappa p_z \bm b} (\bm r_0) \\\
     & = \int \frac{d\varphi_{\bm p}}{2 \pi} (-i)^{m_\ell}\, e^{i m_\ell \varphi_{\bm p}}\, e^{-i \varkappa b \cos \varphi_{\bm p}}\, f_{2p\, m_f, 1s}^{FB} .
     \end{split}
\end{equation}
In the first line of this expression $\psi^{V}_{m_\ell \varkappa p_z \bm b}$ is given by Eq.~\eqref{pw_vortex_wf}, in which the transverse momentum is fixed to $|\bm p_\perp| = \varkappa$ and the longitudinal momentum is $p_z$. In the second line of Eq.~\eqref{amp_pw_v}, we expressed the scattering amplitude of the vortex electron in terms of its plane wave counterpart. The Born approximation for incident vortex electrons~\eqref{amp_pw_v} has been derived by Ruben van Boxem and coauthors in Ref.~\cite{van2015inelastic}.
\subsection{Distorted wave approximation}\label{subsec:dw}
For the conventional (non-vortex) electrons, the distorted wave approximation is well-established~\cite{Chauhan2005} and yields the amplitude
\begin{equation}\label{dwamp_nv}
    f_{2p\, m_f, 1s}^{DW} =  \frac{1}{(2 \pi)^3} \int d^3 \bm r_0\ d^3 \bm r_1 \ F_{\bm p'}^{(-) *} (\bm r_0) \psi_{2p\, m_f}^* (\bm r_1) \frac{1}{| \bm r_0 - \bm r_1|} F_{\bm p}^{(+)} (\bm r_0) \psi_{1s}(\bm r_1).
\end{equation}
In contrast to the first Born theory, the incoming and outgoing electrons are described here by the distorted wave functions $F_{\bm p}^{(+)}$ and $F_{\bm p'}^{(-)}$, given by Eq.~\eqref{dw_wf_eq}.

In order to evaluate the amplitude $f_{2p\, m_f, 1s}^{DW}$ further one can employ the multipole expansion of the electron-electron interaction 
\begin{equation}\label{mp_exp}
    \frac{1}{| \bm r_0 - \bm r_1|} = \sum_{L} \sum_{M} \frac{4 \pi}{2 L +1} Y_{L M}^*( \hat{\bm r}_1) Y_{L M}( \hat{\bm r}_0) \frac{r_< ^{L}}{r_>^{L+1}},
\end{equation}
where $r_< = \min(r_0,\, r_1)$ and $r_> = \max(r_0,\, r_1)$~\cite{fliessbach2012elektrodynamik}. By inserting the expansion~\eqref{mp_exp} into Eq.~\eqref{dwamp_nv} and performing angular integration, we can write the scattering amplitude as
\begin{equation}\label{scatamp_dw_eval}
    f_{2p\, m_f, 1s}^{DW} = \frac{4 \sqrt{2}}{2 \pi} \frac{1}{p p'} (-1)^{-m_f} \sum_{l l' m} Y_{l' m-m_f} (\hat{\bm p}') Y_{l m}^*(\hat{\bm p}) S_{l l' m m_f},
\end{equation}
where the integral over the projectile electron radial components is given by
\begin{equation}
\begin{split}\label{S}
    S_{l l' m m_f} = & \sqrt{\frac{2l+1}{2l'+1}} i^{(l-l')} e^{i(\delta_l + \delta_{l'})}  (l \ 0 \ 1 \ 0 | l' \ 0)  (l\ m \ 1 \ -m_f | l \ m-m_f) \\
    & \int d r_0\ \phi_{l'}^{(-) *} (p', r_0) \phi_l^{(+)} ( p,  r_0) e^{-3/2 r_0} \frac{-64 - 96 r_0 -72 r_0^2 -27 r_0^3 + 64 e^{3/2 r_0}}{81 r_0^2}.
    \end{split}
\end{equation}
For further evaluation of the function $S$ and, hence, of the scattering amplitude~\eqref{scatamp_dw_eval}, knowledge about the radial functions $\phi_l^{(+)} $ and $\phi_{l'}^{(-)}$ is demanded. In general, no analytic expressions for these functions are known and their numerical evaluation will be discussed later in Sec.~\ref{sec:numerics}

We are now prepared to derive the amplitude for the most interesting case, the inelastic scattering of a vortex electron distorted by the Coulomb field of a target. Assuming that the outgoing electron is detected in a non-vortex state, this amplitude reads as
\begin{equation}\label{amp_dw_v}
    \begin{split}
        f_{2p m_f, 1s} ^{DW, V} = & \frac{1}{(2 \pi)^3} \int d^3 \bm r_0\ d^3 \bm r_1 \ F^{(-) *}_{\bm p '} (\bm r_0) \psi_{2p m_f}^* (\bm r_1) \frac{1}{| \bm r_0 - \bm r_1|} F^{V, (+)}_{m_\ell \varkappa p_z \bm b} (\bm r_0) \psi_{1s}(\bm r_1)\\
        = & \frac{4 \sqrt{2}}{2 \pi} \frac{1}{p p'}(-1)^{-m_f} \sum_{l l' m} (-1)^{m_\ell -m} Y_{l' m-m_f} (\hat{\bm p}')    J_{m_\ell-m} (\varkappa b) \Theta_{l m}^* (\theta_{\bm p})   S_{l l' m m_f} .
    \end{split}
\end{equation}
Here we made use of the explicit form of the Coulomb distorted (non-) vortex wave functions~\eqref{dw_wf_eq} and~\eqref{distorted_nv_wf} and performed angular integration similar to that in Eq.~\eqref{dwamp_nv} --~\eqref{scatamp_dw_eval}.
In what follows, we will use this expression to calculate inelastic scattering of vortex electrons by a hydrogen atom and discuss the effect of the Coulomb field distortion. For the latter, we will compare results based on Eq.~\eqref{amp_dw_v} with the first Born predictions~\eqref{amp_pw_v}, which employ the plane wave solution for incoming and outgoing electrons in place of their distorted counterparts $F^{V, (+)}_{m_\ell \varkappa p_z \bm b}$ and $F^{(-)}_{\bm p '}$.
\section{Numerical details}\label{sec:numerics}
As seen from Eqs.~\eqref{scatamp_dw_eval} and~\eqref{amp_dw_v}, calculations of the scattering amplitudes for the distorted (non-) vortex solutions require the knowledge about the radial components $\phi_l^{(+)} $ and $\phi_{l'}^{(-)}$ of the electron moving in the field of a target atom. These radial components are obtained numerically by solving the radial Schrödinger equation
\begin{equation}    \label{eqn:radial}
    \left( \frac{\text{d}^2}{\text{d}r^2_0}  +   p^2 - 2U_d(r_0) - \frac{l(l+1)}{r_0^2}\right)  \phi_{l}(p,r_{0}) = 0 ,
\end{equation}
where $U_d$ is the distortion potential~\cite{Walters1984,Salvat1995,Chauhan2005}. In the present study, this static potential describes the interaction of an electron with a hydrogen atom and is given by
\begin{equation}    \label{eqn:potential}
   U_d(r_0) = -\frac{1}{r_{0}} + \int_0^{\infty} \frac{|R(r_1)|^2}{r_>} r_1^2 \text{d}r_1  ,
\end{equation}
where $R(r_1)$ denotes the radial part of the initial $\ket{1s}$ or final $\ket{2p}$ atomic states. The above integral can be evaluated trivially, yielding $U_d$ as a function of the radial coordinate $r_0$ of the projectile electron. 

The phase shift $\delta_{l}$ in Eq.~\eqref{S} is obtained by solving Eq.~\eqref{eqn:radial} while imposing the following asymptotic boundary conditions~\cite{Walters1984,Salvat1995,Chauhan2005},
\begin{align}
    \phi_l(p, r_{0}) & \xrightarrow{r_0 \rightarrow 0 }  0    \, , \\
    \phi_l(p, r_{0}) & \xrightarrow{r_0 \rightarrow \infty } \frac{1}{\sqrt{p}}\sin \left( pr_{0} - \frac{l\pi}{2} +  \delta_l \right)  \, .
\end{align}
To compute the scattering amplitudes in Eqs.~\eqref{scatamp_dw_eval} and~\eqref{amp_dw_v}, we have taken into account up to 120 partial waves in both the initial and final channels. Acknowledging that the contribution of higher partial waves comes at a higher radial distance, we selected a radial box size of 180 atomic units with a fine radial grid of 4000 points to ensure higher accuracy in our numerical calculations.
\section{Results}\label{sec:results}
\subsection{Distorted vortex electrons: Probability density and phase}\label{sec:results1}
\begin{figure}[h!]
    \centering
    \includegraphics[width=0.7\linewidth]{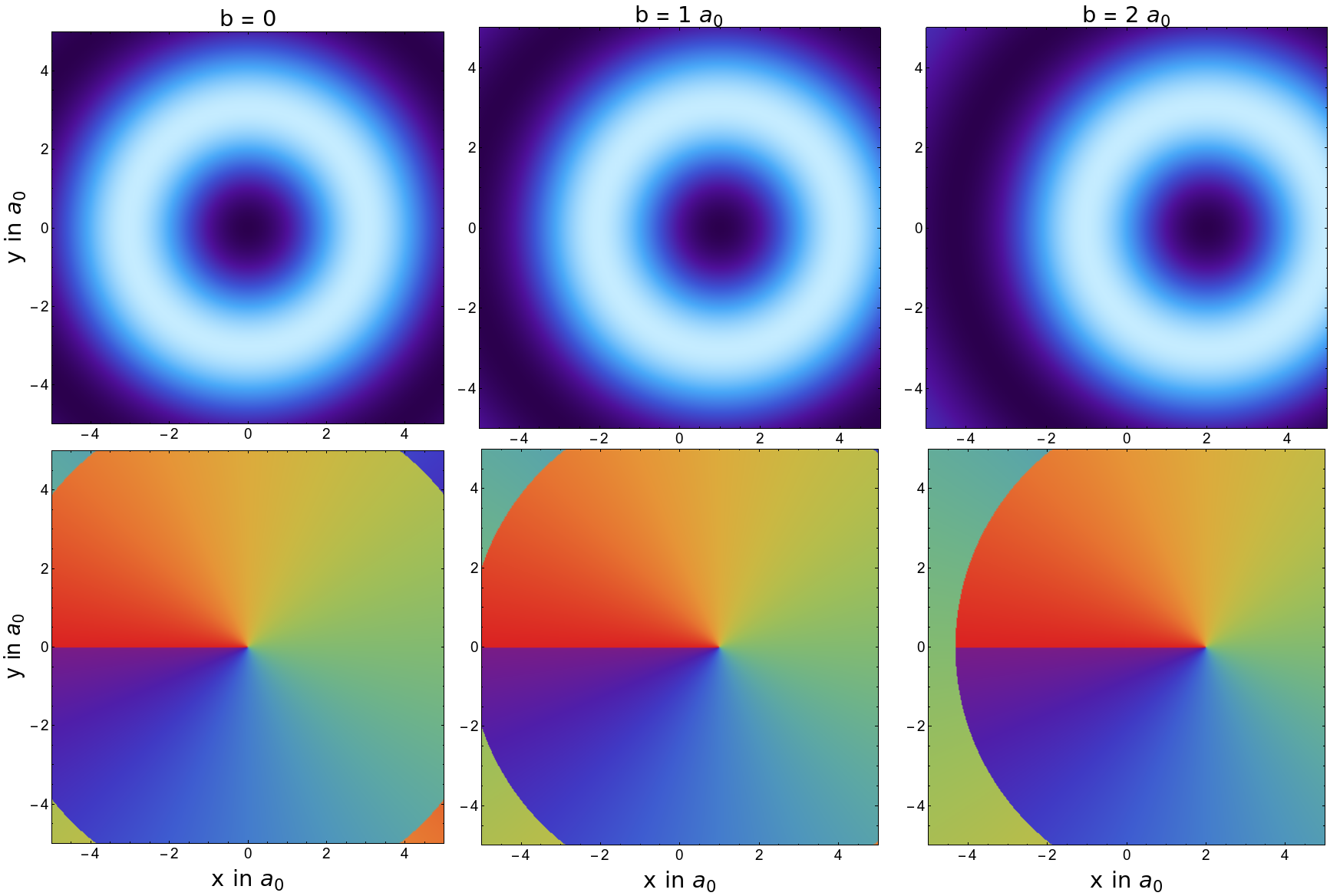}
    \caption{Upper panel: Probability density of a 20 eV vortex electron constructed from plane waves~\eqref{pw_vortex_wf} in the $xy$ plane. The columns show different impact parameters $b$. The opening angle is $\theta_{\bm{p}} = 30^\circ$ and the electron has an orbital angular momentum projection of $+1$. Lower panel: Phase of the electrons described above.}
    \label{probdenpw}
\end{figure}

\begin{figure}
    \centering
    \includegraphics[width=0.7\linewidth]{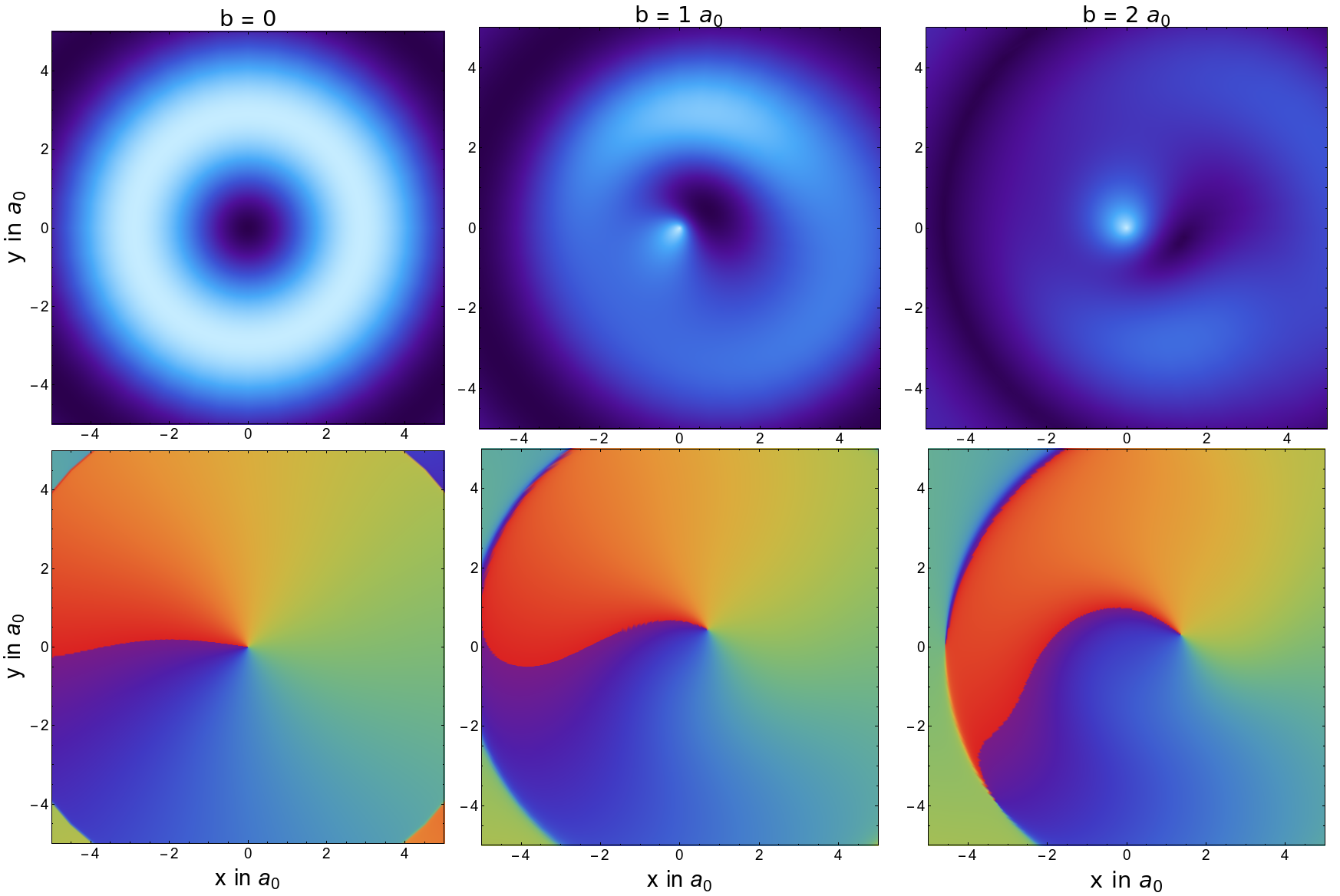}
    \caption{Upper panel: Probability density of a 20 eV vortex electron constructed from distorted waves~\eqref{distorted_nv_wf} in the $xy$ plane. The columns show different impact parameters $b$. The opening angle is $\theta_{\bm{p}} = 30^\circ$ and the electron has an orbital angular momentum projection of $+1$. Lower panel: Phase of the electrons described above.}
    \label{probdendw}
\end{figure}

\begin{figure}
    \centering
    \includegraphics[width=0.7\linewidth]{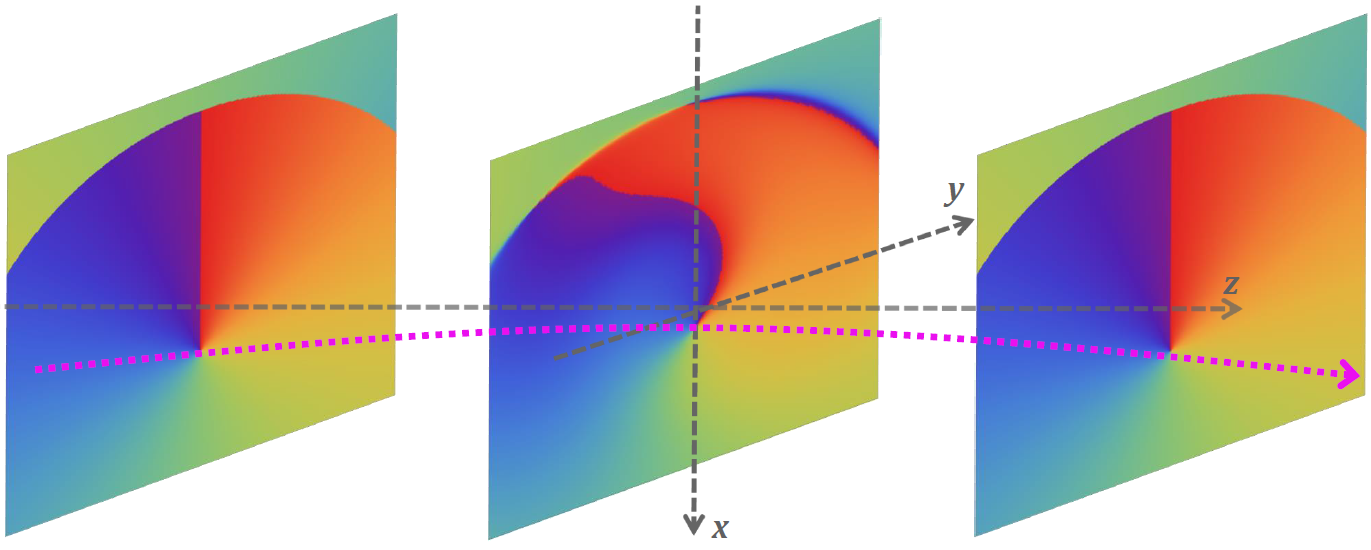}
    \caption{Schematic sketch of the behavior of the phase pattern in the transverse plane of during the propagation of the electron vortex beam along the $z$ axis with (asymptotic) impact parameter $b = 2\, a_0$. The target atom is assumed to be located at the coordinate origin.}
    \label{vortexpos}
\end{figure}

Before addressing the inelastic scattering of vortex electrons, we aim to discuss their properties and pay special attention to the effect of the Coulomb distortion. To achieve this goal, we compare in Figs.~\ref{probdenpw} and~\ref{probdendw} the probability densities 
$|\psi^{V}_{m_\ell \varkappa p_z \bm b} (\bm r)|^2$ and  $| F^{V, (+)}_{m_\ell \varkappa p_z \bm b} (\bm r)|^2$ as well as the phases $\arg (\psi^{V}_{m_\ell \varkappa p_z \bm b} (\bm r))$ and  $\arg( F^{V, (+)}_{m_\ell \varkappa p_z \bm b} (\bm r))$ of vortex electrons moving either in free space or in the presence of the Coulomb field of a target. Calculations are performed for the kinetic energy 20 eV, the opening angle $\theta_{\bm p} = 30^\circ$ and OAM projection $m_\ell=+1$. Moreover, we display the probability densities and phases in the transverse $xy$ plane containing a target atom, i.e. for $z=0$. The three columns of each figure correspond to three various displacements of the vortex line from the center of coordinates along the $x$ axis, represented by $b = 0,\, 1,\, 2\, a_0$. As seen from Fig.~\ref{probdenpw}, shifting the vortex line for the case of vanishing Coulomb interaction merely displaces the ring-shaped probability profile and the winding phase by the corresponding impact parameter $b$. This behavior can be well expected since the wave function $\psi^{V}_{m_\ell \varkappa p_z \bm b} (\bm r)$ describes properties of vortex electrons in free and homogeneous space. Qualitatively different behavior is observed for the vortex electron beam propagating in the field of a target atom. As seen from Fig.~\ref{probdendw}, displacement of the vortex line from the origin of coordinates, i. e., from the Coulomb center, results not only in the translation of the probability density and phase patterns, but also in their significant distortion. 
For instance, as seen from the middle and right columns of Fig.~\ref{probdendw}, the density profile of the vortex electron in the Coulomb field does not exhibit anymore the ``doughnut" shaped probability density, typical for free vortex electrons. This shape is altered by the interaction with the target, which leads in particular to a high probability density spot at the position of the atomic nucleus (placed at the coordinate origin). Furthermore, as shown in the lower line of Fig.~\ref{probdendw}, the Coulomb interaction with the atom strongly influences the phase pattern of the vortex electron beam. Particularly, while the phase front of an electron propagating in free space exhibits a well-known $e^{i m_\ell \varphi_{\bm p}}$ dependence (see Fig.~\ref{probdenpw}), the phase of the distorted vortex electrons depends additionally on the distance to the vortex line. This effect can be attributed to the well-known Coulomb phase which enters in the radial component of the distorted wave function, $\phi_l^{(\pm)} \sim  e^{\pm i( p r - \eta  \ln (2 p r))}$, see Ref.~\cite{eichler2005lectures} for further details. The Coulomb interaction of the target and vortex electron beam also leads to the fact that the vortex line is not placed anymore at $x = b$, but is shifted towards the Coulomb center, see Fig.~\ref{vortexpos}. Our calculations performed for various effective nuclear charges have shown that this displacement from the position, expected for the free vortex electrons, increases with the strength of the Coulomb distortion.

The effect of the Coulomb distortion on the density and phase of the vortex electrons is most pronounced, of course, at the longitudinal coordinate $z=0$ corresponding to the scenario of the closest approach between the electron and the target atom. For the asymptotic domain $z \rightarrow \pm \infty$, where the interaction with the target becomes negligible, the behavior of the vortex electron beam recovers the well-known free space solutions. Specifically, $|\psi^{V}_{m_\ell \varkappa p_z \bm b} (\bm r)|^2$ has an azimuthally symmetric doughnut shape and a $e^{i m_\ell \varphi_{\bm p}}$ helical phase behavior with the vortex line located at $x =b$ if $|z|$ is large.
\subsection{Scattering amplitudes}
\begin{figure}
    \centering
    \includegraphics[width=0.7\linewidth]{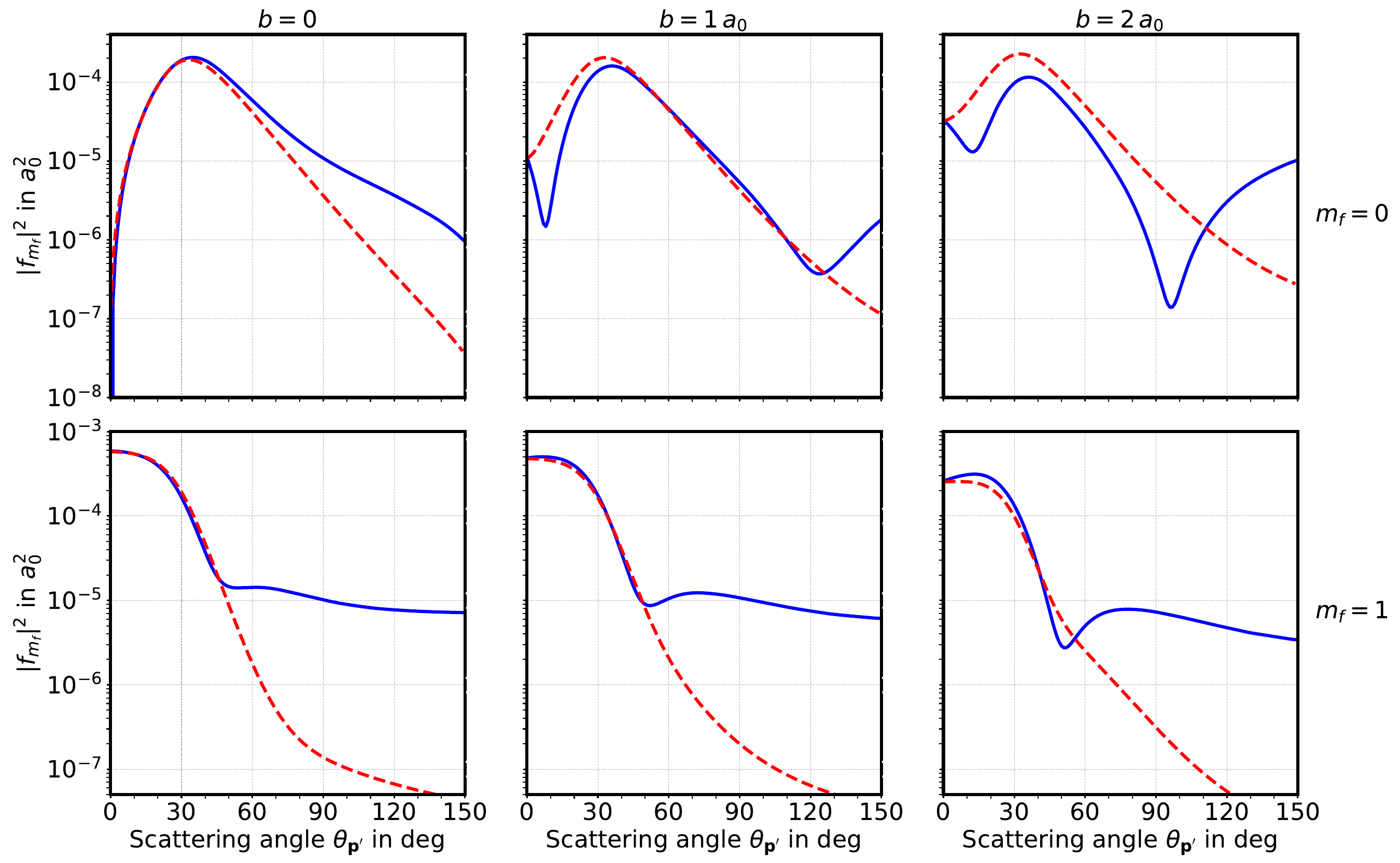}
    \caption{Squared amplitude as a function of the polar scattering angle $\theta_{\bm{p}'}$ for the $1s \rightarrow 2p$ excitation of hydrogen. Predictions of the first Born and distorted wave approximations are displayed by red dashed and blue solid lines, respectively. The columns correspond to the impact parameters $b = 0,\, 1,\, 2\, a_0$. Results for the excitation to the final magnetic substates $\ket{2 p \, m_f=0}$ and $\ket{2 p \, m_f=1}$ are shown in the top and bottom rows. Additionally, the initial vortex electrons have a kinetic energy of 20 eV, opening angle $\theta_{\bm{p}} = 30^\circ$ and OAM projection  $m_\ell = +1$.}
    \label{ampsq_angdep}
\end{figure}

\begin{figure}
    \centering
    \includegraphics[width=0.7\linewidth]{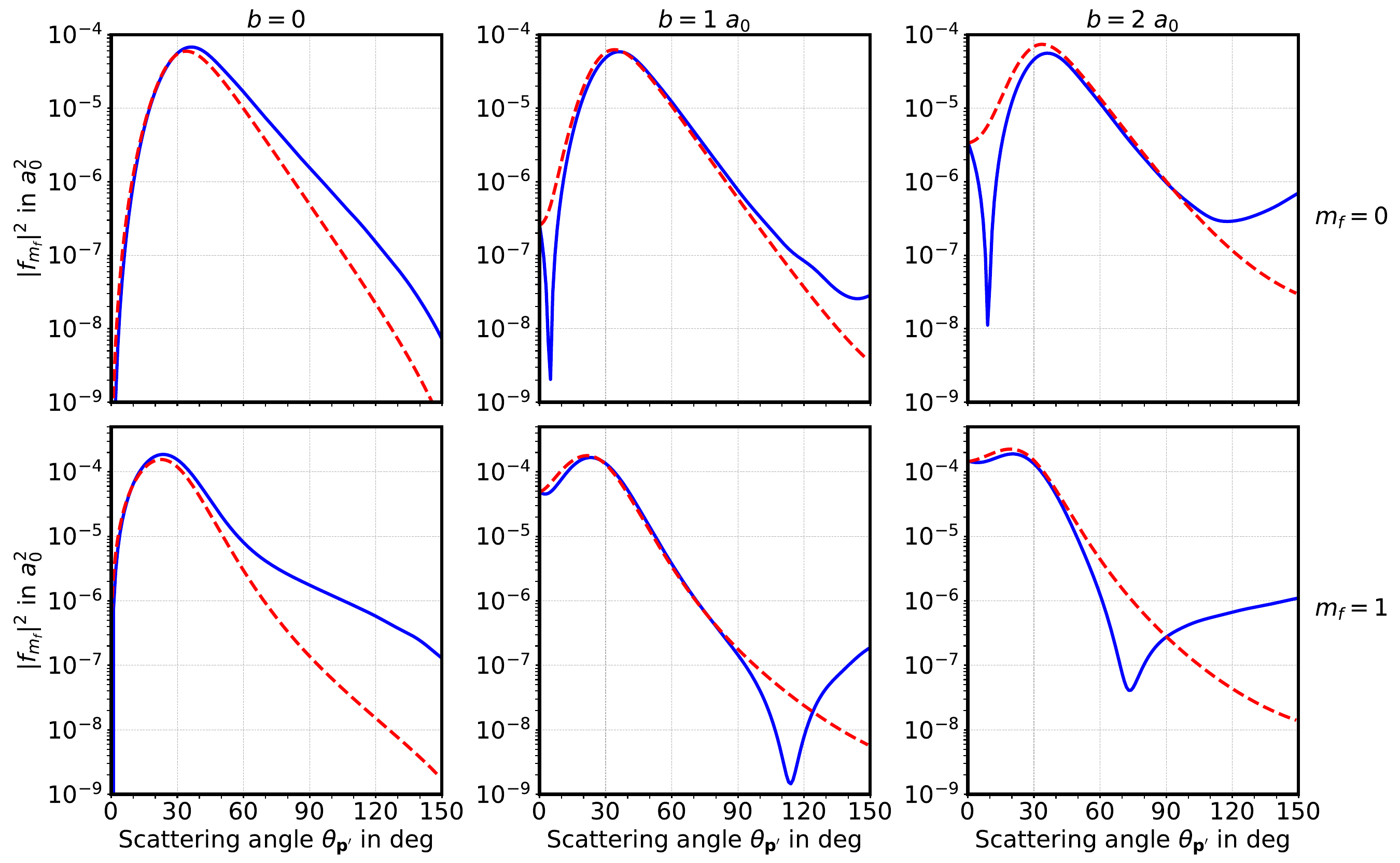}
    \caption{The same as Fig.~\ref{ampsq_angdep}, but for OAM projection $m_\ell=+2$.}
    \label{ampsq_angdepOAM2}
\end{figure}

After discussing the properties of free and distorted vortex electrons, we now turn to the analysis of their inelastic scattering by a hydrogen atom. In Figs.~\ref{ampsq_angdep} and~\ref{ampsq_angdepOAM2}, for example, we display the squared scattering amplitudes that describe excitation of an atom to the $\ket{2 p\, m_f = 0}$ (upper row) and $\ket{2 p \, m_f=1}$ (lower row) magnetic substates. The calculations have been performed for the incident Bessel electrons with kinetic energy 20 eV, opening angle $\theta_{\bm p} = 30^\circ$ and OAM projections $m_\ell = +1$ and $m_\ell = +2$. Moreover, we assumed that the vortex line is asymptotically displaced from the coordinate origin by the impact parameters $b=0$ (left column), $b=1\, a_0$ (middle column) and $b=2\, a_0$ (right column). To investigate the effect of the Coulomb interaction, we show the polar angle dependence for the free (red dashed line) and distorted (blue solid line) vortex electrons. 

As seen from the left column of Figs.~\ref{ampsq_angdep} and~\ref{ampsq_angdepOAM2}, the first Born and distorted wave predictions are in good agreement for the scattering under small angles, $\theta_{\bm p '} \lesssim 30^\circ$, and for the hydrogen atom placed on the vortex line, $b=0$. In particular, both theories forbid strictly forward scattering ($\theta_{\bm p'}= 0$) with the excitation of the $\ket{2 p \, m_f= 0}$ state. Moreover, the $\ket{1s} \rightarrow \ket{2p \, m_f=1}$ transition is also prohibited for the OAM projection $m_\ell = + 2$ but is pronounced for $m_\ell = +1$. This behavior is a clear manifestation of the  selection rule 
\begin{equation}
    m_i +m_\ell = m_f,
\end{equation}
which can be deduced from the amplitudes~\eqref{amp_pw_v} and~\eqref{amp_dw_v} for $b=0$ and $\theta_{\bm p '} = 0$, and, hence, of the conservation of OAM projections. For the excitation of the hydrogen atom from the ground state, $m_i = 0$, the selection rule indicates that strictly forward scattering is allowed only for the case $m_\ell = m_f$.

While for a hydrogen atom placed on the vortex line, the first Born and distorted wave approximations show a good agreement  for $\theta_{\bm p '}\lesssim 30^\circ$, their predictions diverge for larger angles. For example, for $b = 0$ and $\theta_{\bm p '} \gtrsim 50^\circ$, the first Born calculations strongly underestimate the distorted wave results. A similar effect is known for the scattering of non-vortex electrons and is attributed to a more accurate treatment of the interaction with the Coulomb center~\cite{Chauhan2005}.
When an atom is displaced from the vortex line, $b \neq 0$, the effect of the Coulomb interaction on the inelastic scattering process becomes noticeable even at smaller scattering angles. This can be seen in the middle and right columns of Figs.~\ref{ampsq_angdep} and~\ref{ampsq_angdepOAM2}, where deviations between the first Born and distorted wave calculations emerge even for small $\theta_{\bm p'}$'s. Notably, for excitation to the $\ket{2p , m_f=0}$ sublevel, $|f_{m_f=0}^{DW, V}|^2$ exhibits a pronounced minimum at $\theta_{\bm p '} \approx10^\circ$ which does not appear in the first Born approximation. Similar minimum structure can be observed for $|f_{m_f=1}^{DW, V}|^2$ for the scattering of vortex electrons with initial OAM projection $+2$, while the position of the minima are shifted to larger scattering angles. 

\begin{figure}
    \centering
    \includegraphics[width=0.7\linewidth]{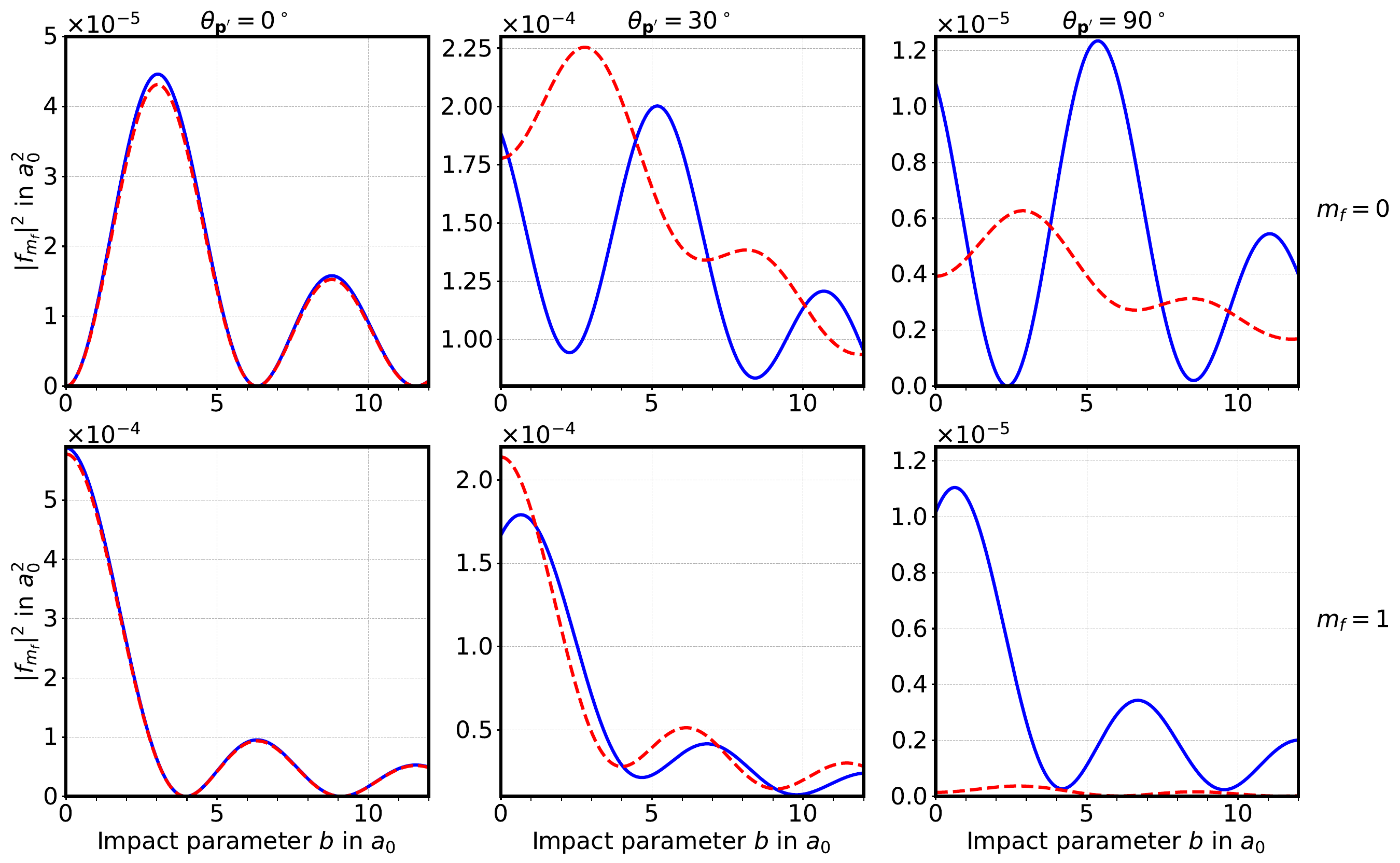}
    \caption{Squared amplitude as a function of the impact parameter $b$ for the $1s \rightarrow 2p$ excitation of hydrogen. Predictions of the first Born and distorted wave approximations are displayed by red dashed and blue solid lines, respectively. The columns correspond to the polar scattering angles $\theta_{\bm{p} '} = 0^\circ,\, 30^\circ, \, 90^\circ$. Results for the excitation to the final magnetic substates $\ket{2 p \, m_f=0}$ and $\ket{2 p \, m_f=1}$ are shown in the top and bottom rows. Additionally, the initial vortex electrons have a kinetic energy of 20 eV, opening angle $\theta_{\bm{p}} = 30^\circ$ and OAM projection  $m_\ell = +1$.}
    \label{diffb}
\end{figure}

To discuss the impact parameter dependence of the scattering process further, we display in Fig.~\ref{diffb} the squared amplitudes $|f_{m_f}^{FB, V}|^2$ (red dashed line) and $|f_{m_f}^{DW, V}|^2$ (blue solid line) as a function of $b$. Analogously to before, calculations have been performed for the kinetic energy 20 eV, opening angle $\theta_{\bm p} = 30^\circ$ and OAM projection $m_\ell = +1$. The results for electron scattering under the angles $\theta_{\bm p '} = 0^\circ, \, 30^\circ, \, 90^\circ$ are displayed in the left, middle and right columns, respectively. As seen from the figure, the first Born and distorted wave results for the strictly forward scattering, $\theta_{\bm p} = 0^\circ$, agree well with each other. In particular, they exhibit oscillations which can be described as 
\begin{equation}
    |f_{m_f}|^2 \sim | J_{m_\ell - m_f} (\varkappa b)|^2.
\end{equation}
This expression is derived within the first Born approximation by taking $\theta_{\bm p '} = 0 $ in Eq.~\eqref{amp_pw_v} and performing analytical integration over $\varphi_{\bm p}$.
At larger $\theta_{\bm p '}$, the first Born and distorted wave predictions may significantly deviate from each other as already mentioned above. For the excitation of the $\ket{2 p \, m_f = 0}$ sublevel, for instance, the $b$-dependent oscillations of $|f_{m_f = 0}^{FB, V}|^2$ and $|f_{m_f = 0}^{DW, V}|^2$ are strongly asynchronous. For $m_f =1$, the oscillations are qualitatively the same, but the Born approximation strongly underestimates the absolute values of the squared amplitude if an electron is scattered under the angle $\theta_{\bm p'} = 90^\circ$.

\begin{figure}[h]
    \centering
    \includegraphics[width=0.8\linewidth]{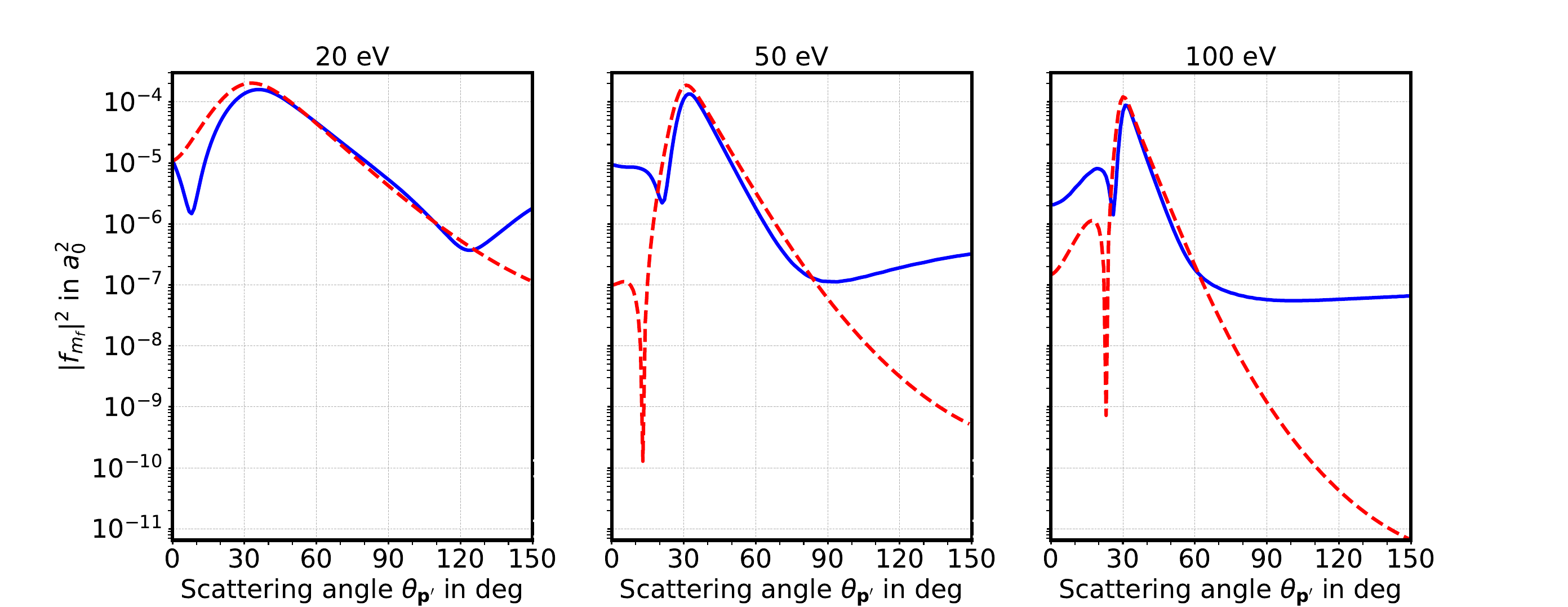}
    \caption{Squared amplitude as a function of the polar scattering angle $\theta_{\bm{p}'}$ for the $\ket{1s} \rightarrow  \ket{2 p \, m_f=0} $ excitation of hydrogen. Predictions of the first Born and distorted wave approximations are displayed by red dashed and blue solid lines, respectively. Calculations have been performed for the kinetic energy 20 eV (left column), 50 eV (middle column) and 100 eV (right column). The initial vortex electrons have the OAM projection of $m_\ell = +1$,  an opening angle $\theta_{\bm p} = 30^\circ$ and impact parameter $b = 1\, a_0$ in all cases.}
    \label{energycomp}
\end{figure}

Until now we have discussed the amplitudes for the scattering of free and distorted vortex electrons for the particular energy of 20 eV. In order to investigate the kinetic energy dependence of the scattering process, we present in Fig.~\ref{energycomp} the squared amplitudes $|f_{m_f=0}^{FB, V}|^2$ and $|f_{m_f=0}^{DW, V}|^2$ for the excitation to the $\ket{2 p\, m_f = 0}$ sublevel of a hydrogen target for $E_{kin} =$ 20, 50, 100 eV. The squared scattering amplitude is displayed as a function of the polar angle $ \theta_{\bm p '}$ and for the fixed impact parameter $b = 1\, a_0$. As seen from the figure, the first Born and distorted wave calculations are in reasonable agreement at intermediate scattering angles across all plots. In contrast, for small and large $\theta_{\bm p '}$, the predictions of the two theories significantly deviate from each other. In  particular, for the high energy regime, $E_{kin} \ge 50$ eV the Born calculations significantly underestimate the distorted wave results for $\theta_{\bm p '} \lesssim 30^\circ$ and $\theta_{\bm p '} \gtrsim 60^\circ $. Both theories, however, predict a remarkable minimum at $\theta_{\bm p '} \approx 20^\circ $ for the higher energies. These minima have been discussed by van Boxem and coauthors in Ref.~\cite{van2015inelastic} and were attributed to the fact that $f_{m_f}^{FB, V} \sim q_z$, where $q_z$ is the longitudinal component of the momentum transfer. It can be easily seen that $q_z = p_z -  p'  \cos (\theta_{\bm p'})$ (see Eq. (32a) in Ref.~\cite{van2015inelastic}) which implies minima in the scattering amplitude at $\theta_{\bm p '} = 13.9^\circ$ and $\theta_{\bm p '} = 24^\circ$ for 50 and 100 eV, respectively.
\section{Conclusion}\label{sec:sum}
In conclusion, we have conducted a theoretical investigation of the inelastic scattering of vortex electrons by a hydrogen target atom, with a particular emphasis on the effects of the Coulomb electron-atom interaction. To explore these effects, we derived expressions for the wave functions and the scattering amplitudes for electron motion, both in free space and in the presence of the Coulomb field of the target atom. Based on the derived formulas, detailed calculations have been performed for the $\ket{1s} \rightarrow \ket{2p\ m_f}$ excitation of the hydrogen atom by vortex electrons with the energies ranging from 20 to 100 eV. Results of these calculations have indicated that the Coulomb interaction significantly affects the wave functions as well as the scattering amplitudes. In particular, the Coulomb distortion leads to bending of the vortex line of the electron beam and a clear modification of its probability density profile; the effect which becomes most pronounced at the closest approach between the electron and the target. The Coulomb distortion also strongly affects the scattering amplitudes, especially at large scattering angles. The remarkable effect of the Coulomb interaction remains also for the rather high kinetic energies of about 100 eV which stresses the importance of a proper account of the Coulomb distortion for the analysis of future experiments on the scattering of vortex electrons.
\section*{Acknowledgements}
J. Sommerfeldt acknowledges support from the German Research Foundation (Deutsche Forschungsgemeinschaft, DFG) under the project 546193616.

\section*{References}
\bibliographystyle{unsrt}
\bibliography{main}

\end{document}